\documentclass[aps,prd,twocolumn,tightenlines,floats,amsmath,superscriptaddress]{revtex4}
\usepackage{graphicx}
\usepackage{amssymb}
\usepackage{amsmath,mathrsfs,verbatim}
\usepackage{times}
\usepackage{animate}
\usepackage{latexsym}
\usepackage[usenames]{color}
\allowdisplaybreaks

\begin{document}

\title{Compact Halo around the Sun Accreted after Dark Matter Dissipative Self Interaction}

\author{Ran Huo}
\email{huor913@gmail.com}

\date{\today}

\begin{abstract}
If dark matter particle can be decelerated due to its dissipative self scattering, except for sinking at the galaxy scale to speed up structure formation, it can also be accreted onto local celestial bodies such as the Sun, forming a compact halo. With some simplified assumptions we develop the Boltzmann equation set based on the partition function of the elliptical orbits, and numerically solve it for the accretion process. We find that the orbited dark matter particles will form a halo around the Sun, with the density profile well fitted to be proportional to $r^{-1.6}$ in a wide range of radius. While around the earth such local halo contribution is always several orders below the galactic component, in a very small region centered around the Sun the sunk dark matter particles can lead to a halo density several orders larger than the background galactic component, in particular in the parameter region of small deceleration speed and large cross section, which is still consistent with current constraints. Such potential dark matter local halo with significantly enhanced density will be a very interesting source for dark matter indirect detection if the corresponding channel exists, we discuss the possibility of the gamma-ray spectrum in the solar direction in some detail as an example.
\end{abstract}

\maketitle

\section{Introduction}

The model that dark matter (DM) has self interaction with a cross section of $\mathcal{O}(1)~\text{cm}^2/\text{g}$~\cite{Spergel:1999mh} is originally suggested as a solution to the core vs.~cusp problem~\cite{Moore:1994yx,Flores:1994gz}, and later possibly to other small scale problems such as the ``too-big-to-fail'' problem~\cite{BoylanKolchin:2011de,BoylanKolchin:2011dk} and the missing satellite problem~\cite{Klypin:1999uc,Moore:1999nt}. Nowadays the improved simulation with the baryonic feedback effect~\cite{Hopkins:2017ycn} and the accumulated observation gradually became consistent with each other, reducing the room for any new physics beyond the vanilla cold DM model such as this self-interacting DM (SIDM) model. But the SIDM model has found other motivations such as interpreting the diversities of the rotation curves~\cite{Kamada:2016euw} and the dwarf galaxies~\cite{Sameie:2019zfo}.

If elastic DM self scattering is acceptable, then the postulation of the DM self scattering being inelastic seems a natural alternative, given the freedom of model building in a non-minimal dark sector. The case that DM up-scatters onto an excited state~\cite{Mohapatra:2001sx,Finkbeiner:2007kk,ArkaniHamed:2008qn,Batell:2009vb,Kaplan:2009de,Loeb:2010gj,Khlopov:2010ik,Fan:2013yva,Fan:2013tia,
Finkbeiner:2014sja,Boddy:2014qxa,Schutz:2014nka,Foot:2014uba,Foot:2016wvj,Boddy:2016bbu,Zhang:2016dck,Blennow:2016gde,Rosenberg:2017qia,Agrawal:2017pnb,DAmico:2017lqj,Buckley:2017ttd,Das:2017fyl,
Foot:2017dgx,Foot:2018qpw,Outmazgine:2018orx,Essig:2018pzq,Choquette:2018lvq,Chang:2018bgx,Alvarez:2019nwt,Huo:2019yhk} and promptly decays back to the original ground state (emitting a dark photon which is not our object) is simple but particularly interesting, in which the DM particle decelerates and consequently varializes to a deeper position in the gravitational well to accelerate the structure formation. On galaxies scales, such process will lead to a dark disk~\cite{Fan:2013yva,Fan:2013tia}, or greatly accelerate the core collapse process of the DM halo. For example, the latter effect is suggested as a solution to the mysterious origin of super massive black hole at a very high redshift~\cite{DAmico:2017lqj,Buckley:2017ttd,Outmazgine:2018orx,Essig:2018pzq,Choquette:2018lvq,Chang:2018bgx,Huo:2019yhk} that seems to violate the Eddington limit of accretion.

Except for the galaxy center or the galactic disk, in a much smaller region~\cite{Agrawal:2017pnb} the known local celestial bodies such as the Sun or the earth also provides a similar local ``deep'' position of the gravitational field to collect the decelerated DM particle. To the knowledge of the author, such possibility has not been explored in detail in literature. A similar DM deceleration followed by gravitational capture process has already been spotted in the context of the traditional DM candidate of the Weakly Interacting Massive Particle (WIMP)~\cite{Gould:1987ir}. However, such process is based on the WIMP nucleon scattering cross section for the initial deceleration, and now get severe constraint~\cite{Aprile:2018dbl}. On the other hand, the DM self scattering cross section can be many orders larger, which actually can overwhelm the shortage of the target DM particle number. A quick estimation with a solar mass of $2\times10^{30}~\text{kg}$ (or $1.2\times10^{57}$ nucleon) gives a scattering/capture rate of $1.4\times10^{16}(\frac{\sigma_{\chi n}}{10^{-46}~\text{cm}^2})(\frac{\rho_\chi}{0.39~\text{GeV}/\text{cm}^3})(\frac{\langle\Delta v\rangle}{240~\text{km}/\text{s}})~\text{s}^{-1}$ for an $80~\text{GeV}$ WIMP, and if we switch to our interested dissipative DM (DDM) model~\footnote{In literature similar models have been named as eXciting DM in~\cite{Finkbeiner:2007kk} or Double-Disk DM in~\cite{Fan:2013yva}.}, similarly the scattering/capture rate for the DDM \emph{in the volume of the Sun} is already $1.2\times10^{14}(\frac{\sigma/m_\chi}{1~\text{cm}^2/\text{g}})(\frac{\rho_\chi}{0.39~\text{GeV}/\text{cm}^3})^2(\frac{\langle\Delta v\rangle}{240~\text{km}/\text{s}})~\text{s}^{-1}$. The initial collisional deceleration can happen far away from the Sun and the DDM is still accreted onto the Sun, so the accretion volume is many orders larger than the volume of the Sun, making the DDM accretion larger than the WIMP accretion. This is a simple demonstration that the DDM accretion can be very significant.

In this paper we develop the first calculation based on the Boltzmann equation for such accretion process. The key to our calculation is the usage of number of DDM particles on each elliptical orbit characterized by its energy and angular momentum as the partition function, which enables robust calculation for the outer part of the resultant halo. Due to the limitation of such approximation, the DDM particles very close to or even inside the Sun cannot be studied directly, but can still be inferred indirectly with potentially large uncertainty. However, the possibility of the existence of a solar DDM halo with number density several orders larger than the galactic component is established, which will be very interesting to the DM indirect detection field.

We will define the model and the approximation schemes in section \ref{sec:2}, and further do some definition and review in section \ref{sec:3}. Then we write down the Boltzmann equation terms for each process in section \ref{sec:4}, and numerically solve it with the results presented and discussed in section \ref{sec:5}. Section \ref{sec:6} provides a brief discussion of the found overdense in the context of DM indirect detection experiments, in particular the \emph{possibility} towards the interpretation of the unexpected gamma-ray spectrum dip. Finally we conclude in section \ref{sec:7}.

\section{The Model and Approximation Scheme}\label{sec:2}

The simplified model is the same as the one used in fluid model for cooling~\cite{Essig:2018pzq} and N-body simulation~\cite{Huo:2019yhk} (see a variant in~\cite{Choquette:2018lvq}). We assume a velocity independent DDM self scattering cross section (actually the ratio of the cross section over the DDM mass) $\sigma/m_\chi$ if and only if in the center-of-mass frame the incoming speed of each DDM particle satisfies $\Delta v/2>v_\text{th}$ (the threshold velocity), then the final state DDM speed in the same frame is $\Delta v'/2=((\Delta v)^2/4-v_\text{th}^2)^{\frac{1}{2}}$. Namely in the up scattering and followed prompt decay, in the center-of-mass frame and for each DDM particle a constant kinetic energy of $m_\chi v_\text{th}^2/2$ is transferred to the dark radiation and dissipated at a constant cross section, once there is such kinetic energy to dissipate. The recoil due to the dark radiation emission is $\mathcal{O}(v_\text{th}^2/c)$ and negligible. We also assume all DM is our interested DDM.

For simplicity, in the calculation we will further ignore
\begin{itemize}
\item Any DDM \emph{elastic} self-scattering, as a key simplification.
\item Any DDM nucleon scattering, which is much smaller as aforementioned.
\item Any DDM annihilation \emph{temporarily}. The annihilation rate will always be too small to modify the local halo accretion noticeably. However, it may be manifested in a DM indirect detection experiment.
\item The size of the Sun to approximate it by a point mass $M$, in order always to use a closed elliptical orbit. But we still use the true density profile of the Sun (BS2005-AGS,~OP model in Ref.~\cite{Bahcall:2004pz}) to calculate the inner gravitational field and the speed of DDM if it is submerged, for the purpose of the scattering kinematics. We can see that the total accreted DM mass is still many orders (18 orders as the extreme case we have ever calculated) below the mass of the Sun finally, so its gravity can be ignored. Moreover, unlike the WIMP case whose orbit will always be partially inside the Sun since the scattering deceleration happens there, in our DDM case the whole orbit can be completely outside the Sun. In such cases the bounded DDM particle initially being on a perfect elliptical orbit is not an approximation but exact. However, later the DDM further scatters to decelerate onto lower orbit, and eventually sink to an orbit mostly or completely submerged into the Sun. Such elliptical orbit approximation enable us the trace the DDM to a region sufficiently close to the Sun.
\item the planets and all other celestial bodies in the solar system while studying the accretion of the Sun.
\end{itemize}
For a microscopic dissipative self-scattering $\vec{u}_1+\vec{u}_2\to\vec{u}'_1+\vec{u}'_2$ event viewed in the solar frame (in this paper we will generally use $\vec{v}$ for a velocity in the Milky Way (MW) rest frame, and $\vec{u}$ for a velocity in the solar frame), we assume
\begin{itemize}
\item The scattering is $s$ wave, so the outgoing particles are always isotropic in angular distribution statistically. Consequently if the final state particles become gravitationally bounded, against the angular momentum $L=m_\chi ru\sin\theta=L_\text{max}\sin\theta$ with a certain speed $u$ ($u'_1$ or $u'_2$ above) at certain radius $r$, the expected number $dN$ is $\propto\sin\theta d\theta$ which is the solid angle, so the differential spectrum can be determined
\begin{align}\label{eq:dNdL}
\frac{dN}{dL}&\propto\frac{\sin\theta d\theta}{d\sin\theta}=\frac{\sin\theta}{\sqrt{1-\sin^2\theta}}=\frac{L}{L_\text{max}\sqrt{L_\text{max}^2-L^2}},\nonumber\\
L_\text{max}&=m_\chi ru=m_\chi r{\textstyle\sqrt{2(\frac{E}{m_\chi}+\frac{GM}{r})}}.
\end{align}
By integration $dL$ over $0$ to $L_\text{max}$ we can check that it is already normalized.

In fact such isotropic assumption holds only in the center-of-mass frame. In the free-free process or the bound-free process which will be discussed soon, the free incoming DDM particle will have a nonzero expectation of velocity in the solar frame, therefore the scattered DM particle has an overall velocity expectation superposed on its isotropic distribution. Later we will calculate the free-free process directly in more detail, taking this subtlety into account. As for the bound-free process, such an overall velocity expectation due to the solar motion itself is not harmful, since the initial angular momentum is also randomly distributed due to the isotropic distribution of collision positions surrounding the Sun.
\item The outgoing DM particles have speed $u'_1=u'_2=\sqrt{\frac{1}{2}(u_1^2+u_2^2)-v_\text{th}^2}$. Exact kinematics gives the final state velocity of $\frac{1}{2}(\vec{u}_1+\vec{u}_2)\pm \hat{n}\sqrt{\frac{1}{4}|\vec{u}_1-\vec{u}_2|^2-v_\text{th}^2}$, where $\hat{n}$ is the random unit direction vector of one outgoing DM particle. According to the above isotropic assumption, \emph{statistically} there will be no overall interference for the magnitude of the velocity, namely $(u'_1)^2=\frac{1}{4}|\vec{u}_1+\vec{u}_2|^2+\frac{1}{4}|\vec{u}_1-\vec{u}_2|^2-v_\text{th}^2=\frac{1}{2}(u_1^2+u_2^2)-v_\text{th}^2=(u'_2)^2$. By a bit misuse of terminology we will refer to the assumption $u'_1=u'_2$ as equipartition. This is of course not held in every microscopic dissipative scattering, but an unbiased estimation with correct energy conservation relation. One can introduce some broadening of the final state energy instead of using the (Dirac) delta energy approximation, and in our following specific calculation for the free-free process we will indeed do so. But for simplicity we will ignore such broadening for the other two processes.
\item In the same spirit, the relative speed for the two colliding particle is statistically $\Delta v=\sqrt{u_1^2+u_2^2}$ in the scattering rate calculation. Again, later in the free-free process we will go beyond such statistical expectation, and calculate the relative speed for dissipative scattering rate in the full phase space.
\end{itemize}

\section{5 Processes and 3 Components}\label{sec:3}

For a specific elliptical orbit with energy $E$ and angular momentum $L$, there are only three possibilities for a microscopic dissipative DM self scattering event, or 5 processes while considering particle filling or removal \emph{on a specific energy state}:
\begin{description}
\item[Free-free process:] two DDM particles are added onto orbits of the same energy (free-free-in) according to our previous equipartition assumption (therefore the two final state DDM particles are either both bounded at orbits with the same energy or both free, which also applies in the following processes); while initially they are both unbounded galactic ones.
\item[Bound-free process:] either a DDM particle is kicked out from the initial orbit (bound-free-out), or two DDM particles are added onto orbits of the same energy (bound-free-in); while in the initial state one of the two DDM particles belongs to a bound orbit and the other is an unbounded galactic one.
\item[Bound-bound process:] either one of the two DDM particles is kicked out from the initial orbit (bound-bound-out), or two DDM particles are added onto orbits of the same energy (bound-bound-in); while initially they are both already bounded at certain orbits and such scattering process only redistributes them onto different orbits.
\end{description}

We will sort the gravitationally bounded DDM particles into the orbited part and the sunk part, that the orbited part can be well described by an aforementioned closed elliptical orbit, with a partition function only depending on $E$ and $L$, while the sunk part cannot due to the failure of such assumption or approximation. Then in our later calculation the above reference of ``bound'' actually means the orbited component. The reason for such categorizing is that we do not have a way to directly study the inner sunk component, due to the failure of the basic tool of elliptical orbit partition function. Our strategy is record the sunk component by the transfer at the lower cutoff of the orbited component.

The third component is the free DDM particle which is not gravitationally bounded. Then the free component, orbited component and sunk component are in the descending order of energy.

\subsection{The Free Galactic Component}
For the free galactic DM particle far away from the Sun, we use the standard halo model, and the partition function is the escape velocity $v_\text{esc}$ truncated Maxwell-Boltzmann distribution
\begin{eqnarray}\label{eq:fg}
f_0(\vec{v})d^3v
&=&\frac{1}{N_\text{esc}(\sqrt{\pi}v_0)^3}e^{-\frac{v^2}{v_0^2}}d^3v~H(v_\text{esc}-v).
\end{eqnarray}
Here $H$ is the heaviside step function. The Maxwell-Boltzmann distribution uses 1D velocity dispersion, which has been switched to the circle velocity $v_0$ at the solar radius and confined on a two-dimensional plane as its estimator. $N_\text{esc}=\frac{4}{\sqrt{\pi}v_0^3}\int^{v_\text{esc}}_0\exp(-\frac{v^2}{v_0^2})v^2dv$ gives the normalization when the escape velocity truncation presents. Numerically for the standard halo model, we use the recent value $v_0=240~\text{km}/\text{s}$~\cite{Reid:2014boa} and $v_\text{esc}=580~\text{km}/\text{s}$~\cite{Monari:2018}, as well as the recent local DM energy density $\rho_\chi=0.39~\text{GeV}/\text{cm}^3$~\cite{Green_2017} \footnote{For historical reason, $v_0=220~\text{km}/\text{s}$ and $\rho_\chi=0.3~\text{GeV}/\text{cm}^3$ are widely used such as in the DM direct detection literature.}.

As seen by an observer at rest to the Sun and therefore moving with velocity $\vec{v}_\odot$ in the MW rest frame, the velocity has relation $\vec{u}+\vec{v}_\odot=\vec{v}$ and the partition function now becomes
\begin{align}\label{eq:fe}
&f_\odot(\vec{u})d^3u
=\frac{2e^{-\frac{u^2+v_\odot^2}{v_0^2}}}{N_\text{esc}\sqrt{\pi}v_0^3}u^2du\int(e^{\frac{2uv_\odot}{v_0^2}\cos\theta}d\cos\theta)\\
=&\left\{\begin{array}{ll}
{\displaystyle\frac{udu}{\sqrt{\pi}N_\text{esc}v_0v_\odot}e^{-\frac{u^2+v_\odot^2}{v_0^2}}2\sinh\frac{2uv_\odot}{v_0^2}}, & u<v_\text{esc}-v_\odot,\smallskip \\
{\displaystyle\frac{udu}{\sqrt{\pi}N_\text{esc}v_0v_\odot}\Big(e^{-\frac{(u-v_\odot)^2}{v_0^2}}-e^{-\frac{v_\text{esc}^2}{v_0^2}}\Big)}, & \begin{array}{c}v_\text{esc}-v_\odot<u\\u<v_\text{esc}+v_\odot,\end{array}\medskip \\
0, & u>v_\text{esc}+v_\odot.
\end{array}\right.\nonumber
\end{align}
In cases that the escape velocity is not considered or effectively taken to be infinity (and $N_\text{esc}\to1$), the first expression reduces to the well-known form found by Gould~\cite{Gould:1987ir}. As the process studied here spans almost $5~\text{Gyr}$ of the whole age of the Sun, after time average the observer velocity $v_\odot$ should go back to $v_0$, while at the moment $v_\odot$ may differ from $v_0$ by a few components.

At last, the gravitational field will also accelerate the free DM particle to distort the phase space distribution. In analogy to Ref.~\cite{Gould:1987ir}, at the position $r$ inside a gravitational field, we can see the free DM particle partition function is
\begin{align}\label{eq:fer}
&f_\odot(u,r)d^3u=\frac{uf_\odot(\sqrt{u^2\hspace{-0.2em}-\hspace{-0.2em}u_\text{esc}^2(r)})d^3\sqrt{u^2\hspace{-0.2em}-\hspace{-0.2em}u_\text{esc}^2(r)}}{\sqrt{u^2-u_\text{esc}^2(r)}}\\
=&\left\{\begin{array}{l}
{\displaystyle\frac{u^2du~e^{-\frac{u^2-u_\text{esc}^2(r)+v_\odot^2}{v_0^2}}}{\sqrt{\pi}N_\text{esc}v_0v_\odot\sqrt{u^2\hspace{-0.2em}-\hspace{-0.2em}u_\text{esc}^2(r)}}
2\sinh\frac{2\sqrt{u^2\hspace{-0.2em}-\hspace{-0.2em}u_\text{esc}^2(r)}v_\odot}{v_0^2}},\smallskip\\
\qquad\qquad \sqrt{u^2-u_\text{esc}^2(r)}<v_\text{esc}-v_\odot, \smallskip\\
{\displaystyle\frac{u^2du}{\sqrt{\pi}N_\text{esc}v_0v_\odot\sqrt{u^2\hspace{-0.2em}-\hspace{-0.2em}u_\text{esc}^2(r)}}
\Big(e^{-\frac{(\sqrt{u^2-u_\text{esc}^2(r)}-v_\odot)^2}{v_0^2}}-e^{-\frac{v_\text{esc}^2}{v_0^2}}\Big)}, \smallskip\\ \qquad\qquad v_\text{esc}-v_\odot<\sqrt{u^2\hspace{-0.2em}-\hspace{-0.2em}u_\text{esc}^2(r)}<v_\text{esc}+v_\odot \smallskip\\
0,~~\qquad \sqrt{u^2-u_\text{esc}^2(r)}>v_\text{esc}+v_\odot.
\end{array}\right.\nonumber
\end{align}
where we have used $\sqrt{u^2\hspace{-0.2em}-\hspace{-0.2em}u_\text{esc}^2(r)}d\sqrt{u^2\hspace{-0.2em}-\hspace{-0.2em}u_\text{esc}^2(r)}=udu$. $u_\text{esc}(r)$ is the escape velocity from the local gravitational field, as mentioned earlier it is determined by the true density profile of the Sun but outside it is simply given by $u_\text{esc}^2(r)=2GM/r$. Such argument shift can be viewed as implementing the Liouville's theorem of phase space, and the factor $u/\sqrt{u^2-u_\text{esc}^2(r)}$ is from the rate calculation within a shell of target collision rate.

\subsection{The Orbited Component}
As for the orbited DM particle, given spherical symmetry, the distribution with the orbital parameters $E$ and $L$ of $\frac{d^2N}{dEdL}$ plays the role of partition function. The energy $E$ and angular momentum $L$ can be related to the elliptical Kepler orbit semi-major axis $a$ and semi-minor axis $b$ by
\begin{equation}\label{eq:ab}
a=\frac{GMm_\chi}{-2E},\qquad b=\frac{L}{\sqrt{-2Em_\chi}}.
\end{equation}
We also define $c=\sqrt{a^2-b^2}$ for short, then the DM particle has a radial range from $a-c$ to $a+c$ (Throughout this paper we will always suppress the $E$ and $L$ dependence of $a$, $b$ and $c$, but it should be understood that the corresponding ones should share the same subscript, \emph{e.g.}, $a_1(E_1)$).

From the two variables $E=\frac{1}{2}m_\chi((\frac{dr}{dt})^2+r^2(\frac{d\theta}{dt})^2)-\frac{GMm_\chi}{r}$ and $L=m_\chi r^2\frac{d\theta}{dt}$ conserved in the orbital motion, we can solve $\frac{dr}{dt}=\sqrt{\frac{2GM}{r}-\frac{-2E}{m_\chi}-\frac{L^2}{r^2m_\chi^2}}$. Now we are interested in the probability of the DM particle being in a radius interval $r\to r+dr$. It is proportional to the time the particle spends on it, or $d~\text{probability}\propto dt(r)=dr/(\frac{dr}{dt})$. Eventually with normalization we find
\begin{equation}
d~\text{probability}=\frac{dr}{\pi a\sqrt{\frac{2a}{r}-1-\frac{b^2}{r^2}}}.
\end{equation}
It can be also rewritten as $rdr/(\pi a\sqrt{c^2-(r-a)^2})$, which implies the region of $a-c<r<a+c$.

\section{The Boltzmann Equation}\label{sec:4}

With the last probability differential we can calculate the bound-free or bound-bound process collision rate. It is useful to at first summarize our constraints in the phase space
\begin{description}
\item[(Statistical) energy conservation] $E_1+E_2-m_\chi v_\text{th}^2=2E$, with the help of the equipartition assumption.
\item[Elliptical orbits radial position] $a_i-c_i<r<a_i+c_i,~~\forall~i$.
\item[Threshold for dissipation] $\Delta v=\sqrt{u_1^2+u_2^2}>2v_\text{th}$.
\item[Free ``really free''] $u>u_\text{esc}(r)$ for free particle.
\item[Cannot escape the MW] $u^2+u_\text{esc}^2(r)<(v_\text{esc}+v_\odot)^2$.
\end{description}

The rate for a general bound-free-out process to remove particles on an $(E,L)$ elliptical orbit reads
\begin{align}\label{eq:bfo}
\text{BFO}(E,L)=&-\frac{d^2N}{dEdL}\int_{a-c}^{a+c}\frac{dr}{\pi a\sqrt{\frac{2a}{r}-1-\frac{b^2}{r^2}}}\nonumber\\
&\times\int_{\text{max}(\sqrt{4v_\text{th}^2-\frac{2E}{m_\chi}-u_\text{esc}^2(r)},u_\text{esc}(r))}^{\sqrt{(v_\text{esc}+v_\odot)^2+u_\text{esc}^2(r)}}f_\odot(u,r)d^3u\nonumber\\
&\times \frac{\rho_\chi}{m_\chi}\sigma\Delta v_\text{bf}(u,E,r).
\end{align}
Here the two lines inside the two integrations are both dimensionless probabilities, to scan over all contributing phase space of the incoming free DDM particles. The $f_\odot(u,r)d^3u$ is given in Eq.~\ref{eq:fer}. The two lower bound for velocity integration are ``threshold for dissipation'' and ``free really free'' conditions respectively, and the upper bound is the ``cannot escape the MW'' condition. And in the last line of we have used the notation for the relative speed of the two initial state DDM particles
\begin{equation}
\Delta v=\left\{\begin{array}{ll}
\sqrt{u^2+\frac{2E}{m_\chi}+u_\text{esc}^2(r)} & \text{bound-free}, \\
\sqrt{\frac{2E_1}{m_\chi}+\frac{2E_2}{m_\chi}+2u_\text{esc}^2(r)} & \text{bound-bound},
\end{array}\right.
\end{equation}
as our last overall approximation point. Then $\frac{\rho_\chi}{m_\chi}\sigma\Delta v$ gives the correct dimension of a rate, for kicking particles off the target orbit.

\smallskip

The rate for a general bound-bound-out process reads
\begin{align}\label{eq:bbo}
\text{BBO}(E,L)=&-\frac{d^2N}{dEdL}\int_{E_\text{min}}^{E_\text{max}}dE_1\int_0^{L_\text{max}(E_1)}dL_1\frac{d^2N}{dE_1dL_1}\nonumber\\
&\hspace{-4em}\times\int_{\text{max}(a-c,a_1-c_1)}^{\text{min}(a+c,a_1+c_1)}\frac{dr~H(\Delta v_\text{bb}(E,E_1,r)-2v_\text{th})}{\pi^2aa_1\sqrt{\frac{2a}{r}-1-\frac{b^2}{r^2}}\sqrt{\frac{2a_1}{r}-1-\frac{b_1^2}{r^2}}}\nonumber\\
&\times\frac{1}{4\pi r^2}\sigma\Delta v_\text{bb}(E,E_1,r).
\end{align}
Now we have to integrate over all possible $(E_1,L_1)$ orbits for the other orbited incoming DDM, which in practical numerical calculation are to be sampled by a finite number of bins. The minimal energy $E_\text{min}$ corresponds to a minimal semi-major axis for which the point mass approximation still effectively holds, and the maximal energy $E_\text{max}$ corresponds to a maximal semi-major axis which should probably be related to whether at such scale the target celestial body can still be viewed as isolated. Here for convenience, in practice we somewhat arbitrarily choose a range with
\begin{equation}\label{eq:Erange}
\frac{E_\text{min}}{m_\chi}=-2^{16}~(\text{km}/\text{s})^2,\qquad \frac{E_\text{max}}{m_\chi}=-2^{3}~(\text{km}/\text{s})^2,
\end{equation}
and sample the (negative value of) energy states by every power of $2$ in this range. Such minimal semi-major axis is $1.455~R_\odot$ and the information extracted based on this orbit will suffer some error, while the maximal semi-major axis is $55.4~\text{AU}$ which should have sufficient coverage of the interested halo outskirt. As for angular momentum, for a certain energy, the largest available angular momentum is achieved at a perfectly circular orbit, which is
\begin{equation}\label{eq:Lmax}
L_\text{max}(E)=m_\chi a(E)\sqrt{-\frac{2E}{m_\chi}}
\end{equation}
where the radius $a(E)$ is given by Eq.~\ref{eq:ab} and $\sqrt{-2E/m_\chi}$ is the circular velocity. In practice we use $13$ bins on the $L/L_\text{max}(E)$ dimension. Moreover, the cut from the ``threshold for dissipation'' condition is expressed in heaviside step function $H$ for convenience, and there are also double ``elliptical orbits radial position'' conditions. The change compared with the first line of Eq.~\ref{eq:bfo} can be understood as the number density from an $(E_1,L_1)$ elliptical orbit being $\frac{Ndr}{\pi a_1\sqrt{2a_1/r-1-b_1^2/r^2}}/(4\pi r^2dr)$, namely the number of DDM particles divided by the volume, with the $dr$s canceled.

\smallskip

The bound-free-in process rate to add DDM particles on an $(E,L)$ elliptical orbit reads
\begin{align}\label{eq:bfi}
\text{BFI}(E,L)=2&\int_{E_\text{min}}^{E_\text{max}}dE_1\int_0^{L_\text{max}(E_1)}dL_1\frac{d^2N}{dE_1dL_1}\nonumber\\
\times&\int_{\text{max}(a-c,a_1-c_1)}^{\text{min}(a+c,a_1+c_1)}\frac{dr}{\pi a_1\sqrt{\frac{2a_1}{r}-1-\frac{b_1^2}{r^2}}}\nonumber\\
\times&\int_{\text{max}(\sqrt{4v_\text{th}^2-\frac{2E}{m}-u_\text{esc}^2(r)},u_\text{esc}(r))}^{\sqrt{(v_\text{esc}+v_\odot)^2+u_\text{esc}^2(r)}} f_\odot(u,r)d^3u\nonumber\\
\times&~\frac{\rho_\chi}{m_\chi}\sigma\Delta v_\text{bf}(u,E_1,r)\nonumber\\
&\hspace{-6em}\times{\textstyle\delta(\frac{m_\chi}{2}(u^2\hspace{-0.2em}-\hspace{-0.2em}u_\text{esc}^2(r)
\hspace{-0.2em}-\hspace{-0.2em}2v_\text{th}^2)\hspace{-0.2em}+\hspace{-0.2em}E_1\hspace{-0.2em}-\hspace{-0.2em}2E)}~\frac{dN}{dL}(E,L,r).
\end{align}
The factor $2$ in front corresponds to that two DDM particles are added (the equipartition assumption). The first four lines are quite similar in structure to the previous ones. The new fifth lines contains the differential spectrum to bring the contribution onto the desired $(E,L)$ orbit. For energy the Dirac delta function is our ``statistical energy conservation'' condition, and for angular momentum the $dN/dL$ is given by Eq.~\ref{eq:dNdL}. In terms of $a$ and $b$ we can rewrite the latter as
\begin{equation}
\frac{dN}{dL}=\frac{1}{\big(m_\chi\frac{r^2}{b}\sqrt{\frac{GM}{a}}\big)\sqrt{\frac{2a}{r}-1}\sqrt{\frac{2a}{r}-1-\frac{b^2}{r^2}}},
\end{equation}
and the last factor on the denominator implies a similar radial position cutoff. As for the former delta function, we can trivially integrate it out with the $du$ from the third line.

\smallskip

Quite similarly, the bound-bound-in process rate reads
\begin{align}\label{eq:bbi}
\text{BBI}(E,L)=&\int_{E_\text{min}}^{E_\text{max}}dE_1\int_0^{L_\text{max}(E_1)}dL_1\frac{d^2N}{dE_1dL_1}\nonumber\\
\times&\int_{E_\text{min}}^{E_\text{max}}dE_2\int_0^{L_\text{max}(E_2)}dL_2\frac{d^2N}{dE_2dL_2}\nonumber\\
&\hspace{-6em}\times\int_{\text{max}(a-c,a_1-c_1,a_2-c_2)}^{\text{min}(a+c,a_1+c_1,a_2+c_2)}\frac{dr~H(\Delta v_\text{bb}(E_1,E_2,r)-2v_\text{th})}{\sqrt{\frac{2a_1}{r}-1-\frac{b_1^2}{r^2}}\sqrt{\frac{2a_2}{r}-1-\frac{b_2^2}{r^2}}}\nonumber\\
\times&\frac{1}{\pi^2a_1a_2~4\pi r^2}\sigma\Delta v_\text{bb}(E_1,E_2,r)\nonumber\\
\times&{\textstyle\delta(E_1\hspace{-0.2em}+\hspace{-0.2em}E_2\hspace{-0.2em}-\hspace{-0.2em}m_\chi v_\text{th}^2\hspace{-0.2em}-\hspace{-0.2em}2E)}~\frac{dN}{dL}(E,L,r).
\end{align}
Two DDM particles are added but the $(E_i,L_i)$ space has been doubly counted, so the overall factor is $1$. In real calculation the delta function of ``statistical energy conservation'' is also trivially integrated out by $dE_1$ or $dE_2$.

\smallskip

At last we will work out the free-free-in term with more care. Since this process has no bounded/orbited particle in the initial state, the velocities can be treated exactly with full respection to their directional information, and later we can see that this will be the most significant channel. Unlike the above one we do not always accomplish the integration over the angle $\theta_i$ of each relative velocity $i=1,2$ with the observer $\vec{v}_\odot$, then Eq.~\ref{eq:fer} becomes
\begin{align}\label{eq:fer2}
f_\odot(\vec{u}_i,r)d^3u_i=&\frac{u_i^2du_i\sin\theta_id\theta_id\phi_i}{N_\text{esc}(\sqrt{\pi}v_0)^3}\nonumber\\
\times&e^{-\frac{u_i^2-u_\text{esc}^2(r)+v_\odot^2-2\sqrt{u_i^2-u_\text{esc}^2(r)}v_\odot\cos\theta_i}{v_0^2}}.
\end{align}
Here the $\theta_i$ angle should be asymptotically defined at infinity for the local gravitational field, but we implicitly approximate it to be the local one. Without loss of generality we choose the DDM particle 1 to be at azimuth $\phi=0$, then the general partition function multiplication have $5$ variables $u_1$, $u_2$, $\theta_1$, $\theta_2$ and $\phi$. In order to determine the final state energy, we write down the exact center-of-mass and the relative speeds
\begin{align}\label{eq:u1u2}
u_c=\frac{1}{2}&\sqrt{u_1^2\hspace{-0.2em}+\hspace{-0.2em}u_2^2\hspace{-0.2em}+\hspace{-0.2em}2u_1u_2(\cos\theta_1\hspace{-0.2em}\cos\theta_2\hspace{-0.2em}+\hspace{-0.2em}
\sin\theta_1\hspace{-0.2em}\sin\theta_2\hspace{-0.2em}\cos\phi)},\nonumber\\
\Delta v_\text{ff}=&\sqrt{u_1^2\hspace{-0.2em}+\hspace{-0.2em}u_2^2\hspace{-0.2em}-\hspace{-0.2em}2u_1u_2(\cos\theta_1\hspace{-0.2em}\cos\theta_2\hspace{-0.2em}+\hspace{-0.2em}
\sin\theta_1\hspace{-0.2em}\sin\theta_2\hspace{-0.2em}\cos\phi)},\nonumber\\
\Delta v'_\text{ff}=&\sqrt{(\Delta v_\text{ff})^2-4v_\text{th}^2}.
\end{align}
After the dissipative scattering denoting the angle between $\vec{u}_c$ and $\Delta\vec{v}'$ as $\alpha$, then the probability at such angle is proportional to $\sin\alpha d\alpha$, and the kinetic energy is $\frac{m}{2}(u_c^2+\frac{1}{4}\Delta v_\text{ff}'^2+u_c\Delta v_\text{ff}'\cos\alpha)$ so the energy differential is $dE'=\frac{1}{2}mv_c\Delta v'_\text{ff}d\cos\alpha$. Eventually $d~\text{probability}/dE'\propto d\cos\alpha/d\cos\alpha=$constant, namely the final state energy is evenly distributed in an interval of $mu_c\Delta v_\text{ff}'$. With such broadening of the final state energy distribution, the ``statistical energy conservation'' condition can be better replaced by the
\begin{description}
\item[Real spectrum energy conservation] $u_c^2+\frac{1}{4}\Delta v_\text{ff}'^2-u_c\Delta v_\text{ff}'<\frac{2E}{m_\chi}+u_\text{esc}^2(r)<u_c^2+\frac{1}{4}\Delta v_\text{ff}'^2+u_c\Delta v_\text{ff}'$, that at some $\alpha$ value such final state energy is achieved.
\end{description}
Eventually the free-free-in rate reads
\begin{align}\label{eq:ffi}
\text{FFI}(E,L)=&\int_{a-c}^{a+c} dr~4\pi r^2\nonumber\\
&\hspace{-6em}\times\iint\iiint_{u_\text{esc}(r)}^{\sqrt{(v_\text{esc}+v_\odot)^2+u_\text{esc}^2(r)}}f_\odot(\vec{u}_1,r)d^3u_1f_\odot(\vec{u}_2,r)d^3u_2\nonumber\\
\times&\frac{\rho_\chi^2}{m_\chi^2}\sigma\Delta v_\text{ff}~H(\Delta v_\text{ff}-2v_\text{th})\nonumber\\
\times&{\textstyle H(2E-m_\chi(u_c^2\hspace{-0.2em}-\hspace{-0.2em}u_\text{esc}^2(r)\hspace{-0.2em}+\hspace{-0.2em}\frac{1}{4}\Delta v'^2\hspace{-0.2em}-\hspace{-0.2em}u_c\Delta v'))}\nonumber\\
\times&{\textstyle H(m_\chi(u_c^2\hspace{-0.2em}-\hspace{-0.2em}u_\text{esc}^2(r)\hspace{-0.2em}+\hspace{-0.2em}\frac{1}{4}\Delta v'^2\hspace{-0.2em}+\hspace{-0.2em}u_c\Delta v')-2E)}\nonumber\\
\times&\frac{1}{m_\chi u_c\Delta v_\text{ff}'}~\frac{dN}{dL}(E,L,r),
\end{align}
with the functions $f_\odot(\vec{u}_i,r)d^3u_i$, $u_c$, $\Delta v_\text{ff}$ and $\Delta v_\text{ff}'$ given just above in Eq.~\ref{eq:fer2} to \ref{eq:u1u2}. Again two DDM particles are added but the free particle phase space has been doubly counted, so the overall factor is $1$. In the second line the integration over the direction of $\phi_1$ which is chosen should give $2\pi$. We use the Monte Carlo integrator \texttt{vegas} to explore the whole phase space spanned by the $5$ velocity variables as well as the radius $r$ in the local gravitational field, with respect to all the cuts.

Eventually with all terms at hand, we can write down the Boltzmann equation for the partition function $\frac{d^2N}{dEdL}$
\begin{equation}\label{eq:Boltzmann}
\frac{d}{dt}\Big(\frac{d^2N}{dEdL}\Big)=\big(\text{FFI}+\text{BFO}+\text{BFI}+\text{BBO}+\text{BBI}\big)(E,L).
\end{equation}

\section{Results}\label{sec:5}

Before we show the sample results, we can see that the energy density has no dependence on the DDM mass $m_\chi$, given that we actually use the value of the ratio $\sigma/m_\chi$ for the DDM self scattering cross section. In fact when multiplying the Boltzmann Eq.~\ref{eq:Boltzmann} by $m_\chi$ on both sides, in each term all the $m_\chi$ factors originally in $\rho_\chi/m_\chi$ can be absorbed either into the partition function $\frac{d^2(m_\chi N)}{dEdL}\equiv\frac{d^2m}{dEdL}$ or into the cross section $\sigma/m_\chi$.

\subsection{The Rates}
\begin{figure}[t]
\includegraphics[scale=0.45]{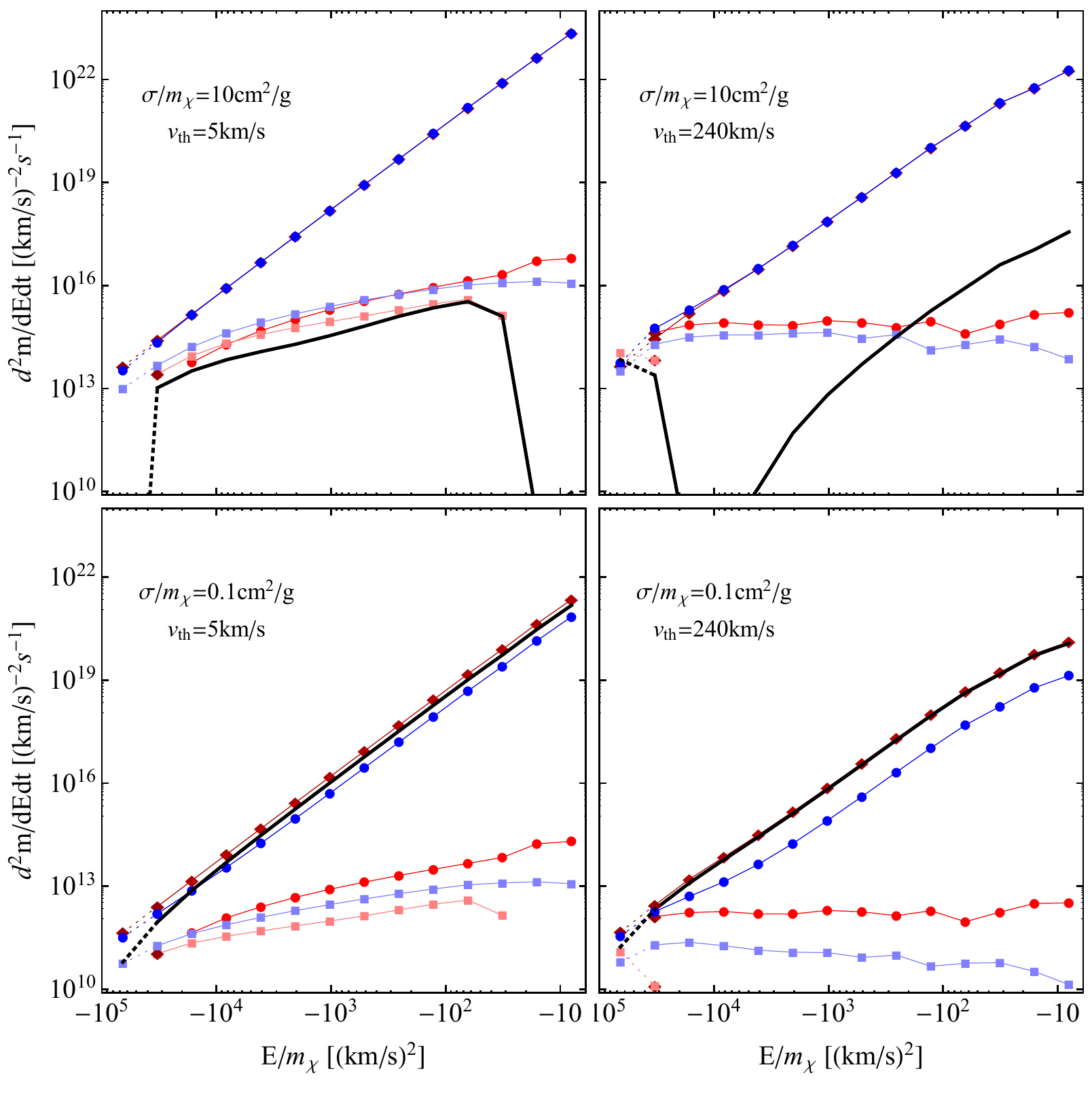}
\caption{The rate at present time ($t=1.5\times10^{17}~\text{s}=4.7~\text{Gyr}$) for 5 processes after integration over angular momentum, for a combination of a small ($\sigma/m=0.1~\text{cm}^2/\text{g}$) or a large ($10~\text{cm}^2/\text{g}$) cross section with a small ($v_\text{th}=5~\text{km}/\text{s}$) or a large ($240~\text{km}/\text{s}$) threshold speed. Red dots and red connecting curves are for all ``in'' processes and blue ones are for all ``out'' processes for a specific energy respectively; and (filled) diamond, circle and square are for free-free process, bound-free processes and bound-bound processes respectively. The black curves denote the net partition function change rate, which are not necessarily positive. The lowest energy bin is close to the sunk component and can be significantly corrected, so we use dotted connecting curves to indicate the possible errors.}
\label{fig:rateAll}
\end{figure}
In Fig.~\ref{fig:rateAll} we first show the sample rates for all the five processes. Except for the constant free-free-in rate, the other four rates depend on at least one partition function of an elliptical orbit so are time dependent, here we choose to show them at the present time of the history of the Sun, which is roughly $t=1.5\times10^{17}~\text{s}$. The energy range is given in Eq.~\ref{eq:Erange}; and for the angular momentum range which we have integrated over, we in fact sample them by a set of angular momentum values $L$ which satisfy $0<L/L_\text{max}(E)<1$.

We can see that very generally in a large $E$ range (or large radius range), the free-free-in rate (red filled diamond) is the largest, in particular for an energy state not very deep in the gravitational potential well, or at the outer part of the halo. But this large rate tends to get balanced by the bound-free-out (blue filled circle) rate, and for a large cross section such cancellation is really good. Later we will see that an equilibrium configuration should have been achieved, so the net partition function after cancellation are many orders smaller. The remaining bound-bound-out rate, bound-free-in rate and bound-bound-in rate are much smaller and somewhat comparable themselves, and their dependence on the depth in the gravitational potential well or radial position is much milder. The bound-free-in rate is much smaller than the bound-free-out process because the latter also contains the events that the final state DDM particles are kicked out of the halo by the energetic incoming DDM particles, and this process is actually dominant. For small $v_\text{th}$  the scattering is almost elastic and the partition function change due to scattering should be small, then the bound-bound-out process should be balanced by the bound-bound-in process, and we can see that it is indeed the case within numerical precisions. But the bound-bound-in rate at a shallow position of the gravitational field is cutoff, since the target energy state must be at least deeper than $-v_\text{th}^2/2+E_\text{max}/m_\chi$. While for a $5~\text{km}/\text{s}$ threshold speed such cutoff only affect the rightmost 3 bins in energy, for a $240~\text{km}/\text{s}$ threshold speed the effect can be very deep. On the other hand, due to the low cutoff energy $E_\text{min}/m_\chi$ we expect the leftmost rate to suffer some error, so we use dotted curves to indicate the possible errors.

\subsection{The Orbit Component and the Outer Halo}
With the numerical solution to the coupled Boltzmann equation set, we can calculate the halo profile from the contribution of the orbited component
\begin{align}
&\rho(r)=\\
&\int_{E_\text{min}}^{E_\text{max}}dE\int_0^{L_\text{max}(E)}dL\frac{d^2m}{dEdL}\frac{H(r-a+c)~H(a+c-r)}{\pi a\sqrt{\frac{2a}{r}-1-\frac{b^2}{r^2}}~4\pi r^2}.\nonumber
\end{align}

\begin{figure*}[ht]
\includegraphics[scale=0.5]{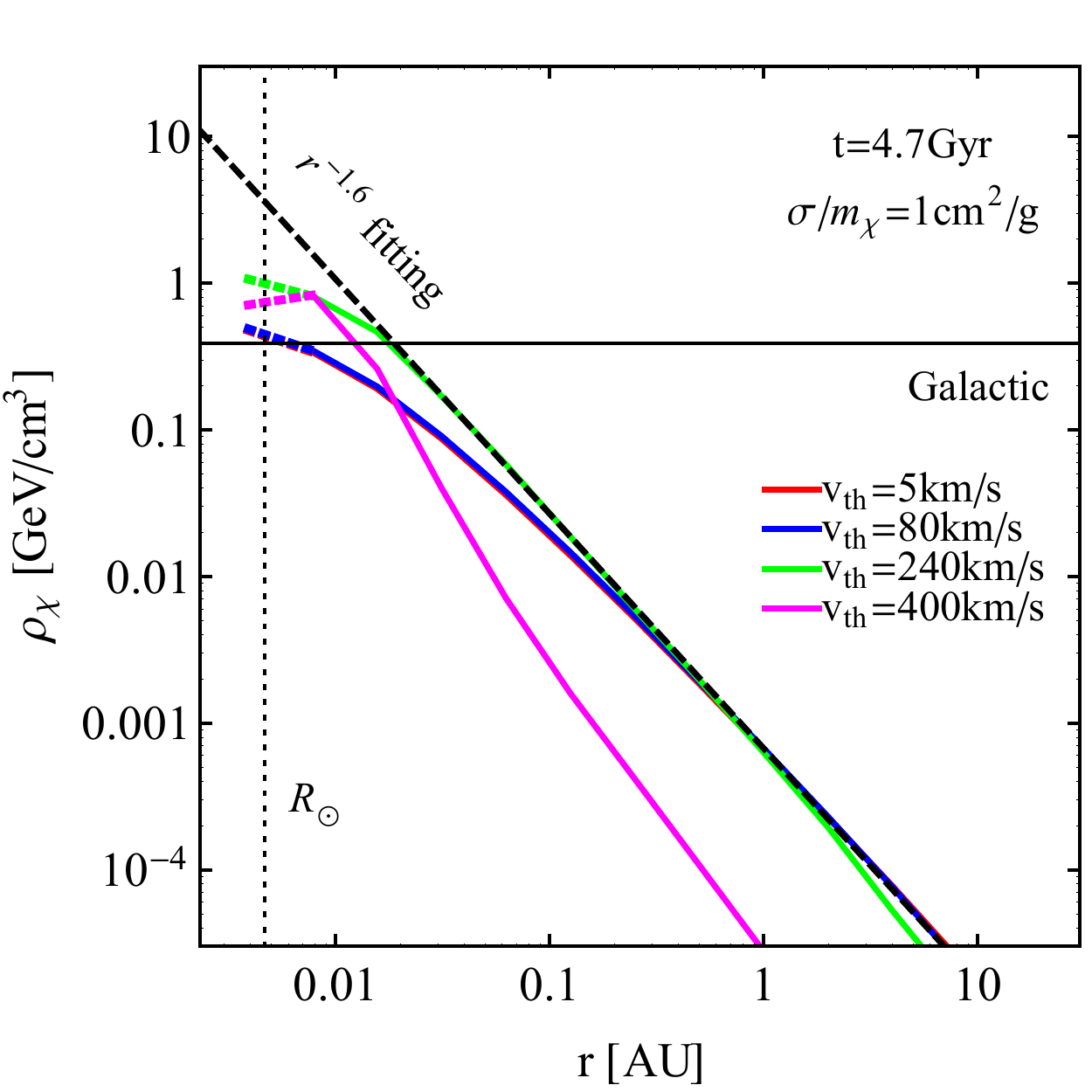}\qquad
\includegraphics[scale=0.5]{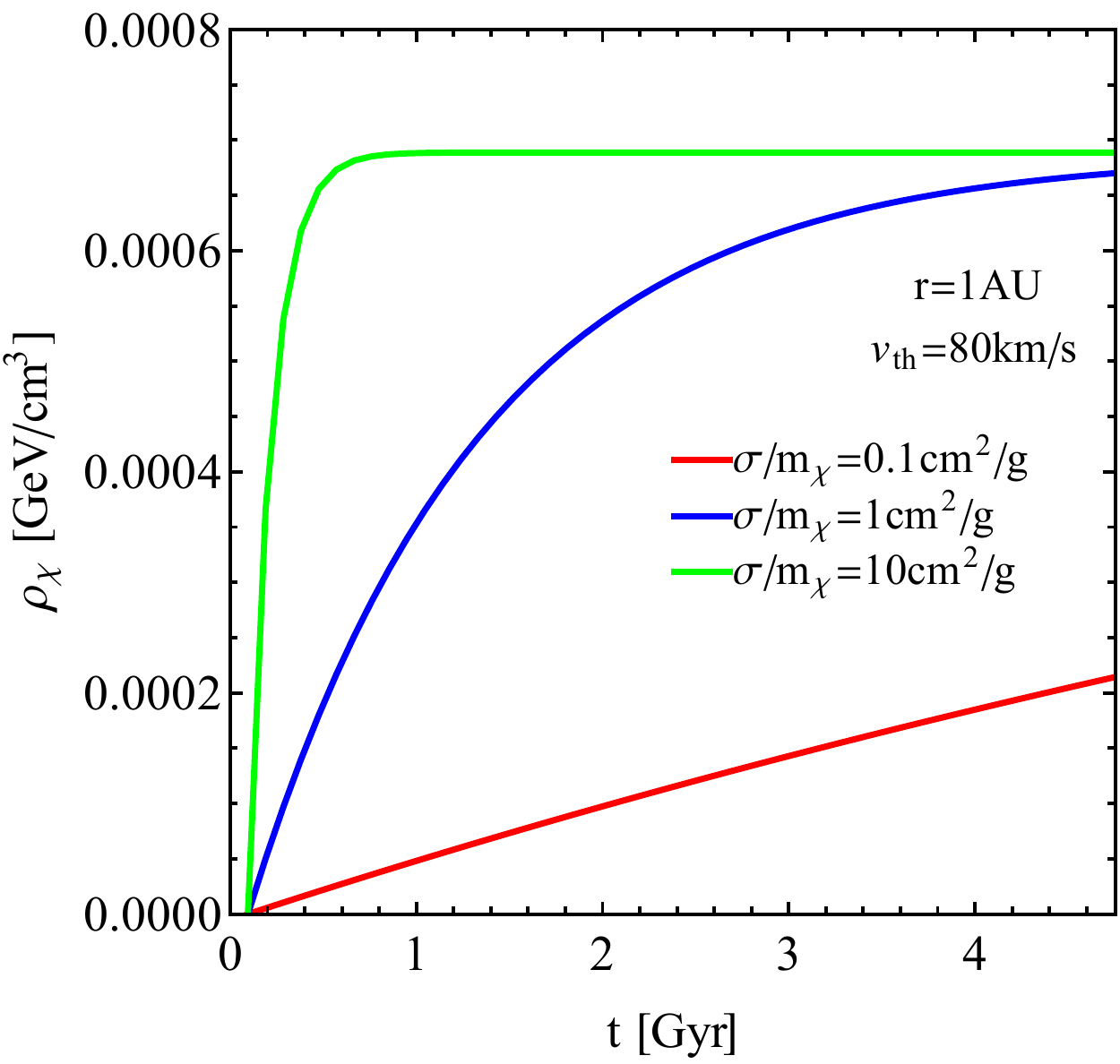}
\caption{Sample spatial and temporal slices of the halo around the Sun, \emph{using the orbited component only}. In the left panel we also fix the cross section, and in the right panel we also fix the threshold velocity. Note that on the radial slice the sunk component can greatly further enhance the DDM density, but this is not included here, so the profile only from the orbited component in the corresponding region is plotted by dotted curves. The evolution of the density show clear equilibrium configuration and the tendency to approach such equilibrium at different speed.}
\label{fig:forRforT}
\end{figure*}

We show the resultant halo by two slices in Fig.~\ref{fig:forRforT}. In the left panel of the radial slice, we have fixed the time as well as the cross section, and choose four representative threshold speeds spanning a wide range of interest. We can see that in a very large radial range the halo can be well fitted by a power law $\rho\propto r^n$ with $n\sim-1.6$ for $v_\text{th}\lesssim300~\text{km}/\text{s}$. In this region the dependence of the halo on the threshold speed is very weak, for example, the $v_\text{th}=5~\text{km}/\text{s}$ curve overlaps with the $v_\text{th}=80~\text{km}/\text{s}$ curve very well. As $v_\text{th}$ further increases the halo becomes more cuspy. The halo density at the position of the earth is about $2.5$ orders below the galactic component, so no effect should be seen such as in the DM direct detection experiment. But at a radius of $r\sim0.01-0.03~\text{AU}$ the halo density increases to equal to the galactic component, and inside the local halo one is even higher. However, later we will see that the sunk component will give an even much more enhanced contribution.

In the right panel of the temporal slice, we have fixed the radius as well as the $v_\text{th}$, and choose three representative cross sections to see the comparison of the accretion speed. We can see that there is an equilibrium configuration, and all the densities corresponding to different cross sections tend to approach such equilibrium configuration upon accumulations. If the cross section is large the current density will be very close to the equilibrium one, and if the cross section is small the current density can still be quite far from the equilibrium one. At last, we can see that the equilibrium configuration will not be far away from the one calculated with $\sigma/m_\chi=1~\text{cm}^2/\text{g}$, namely the solar halo contribution to the terrestrial DM density will always be at least about $2.5$ orders below the galactic contribution and unimportant.

\subsection{The Sunk Component}
As mentioned before, while we can directly sample the orbited component by a set of discretized bins for the differential equation in the $(E,L)$ space, we cannot do so directly for the sunk component due to the intrinsic failure of the elliptical orbit approximation. Such component has to be counted by the transition at the lower boundary $E_\text{min}$ of our energy range. Here we will naively estimate the sunk component as the particles sunk in the bound-bound processes, in particular
\begin{align}\label{eq:sunk}
&N_\text{sunk}=\\
&-\int dt\int_{E_\text{min}}^{E_\text{max}}dE\int_0^{L_\text{max}(E)}dL(\text{BBO}(E,L)+\text{BBI}(E,L)).\nonumber
\end{align}

This estimation goes as follows. In our previous definition of 5 processes, we do not specify the final state being free or bound (orbit) or sunk for an ``out'' process, which suffices the removal rate calculation for a specific orbit. If we always specify the final state, the previous 3 possibilities in the initial state categorizing should by enumeration be expanded to the free-free to bound-bound processes, the bound-bound to bound-bound and bound-bound to sunk-sunk processes, as well as the bound-free to free-free and bound-free to bound-bound processes. We have actually ignored the free-free to sunk-sunk process since it need to go beyond the equipartition assumption and use the broadened spectrum (the real spectrum energy conservation) of the final state kinetic energy, although from Fig.~\ref{fig:rateAll} we can see that by doing this a small rate can be really extrapolated. Then we can see that the previous BBO process corresponds to the sum of the bound-bound to bound-bound and bound-bound to sunk-sunk processes, subtracting the bound-bound to bound-bound (BBI) process is indeed the only channel contributing to the sunk component.

\begin{figure}[ht]
\includegraphics[scale=0.5]{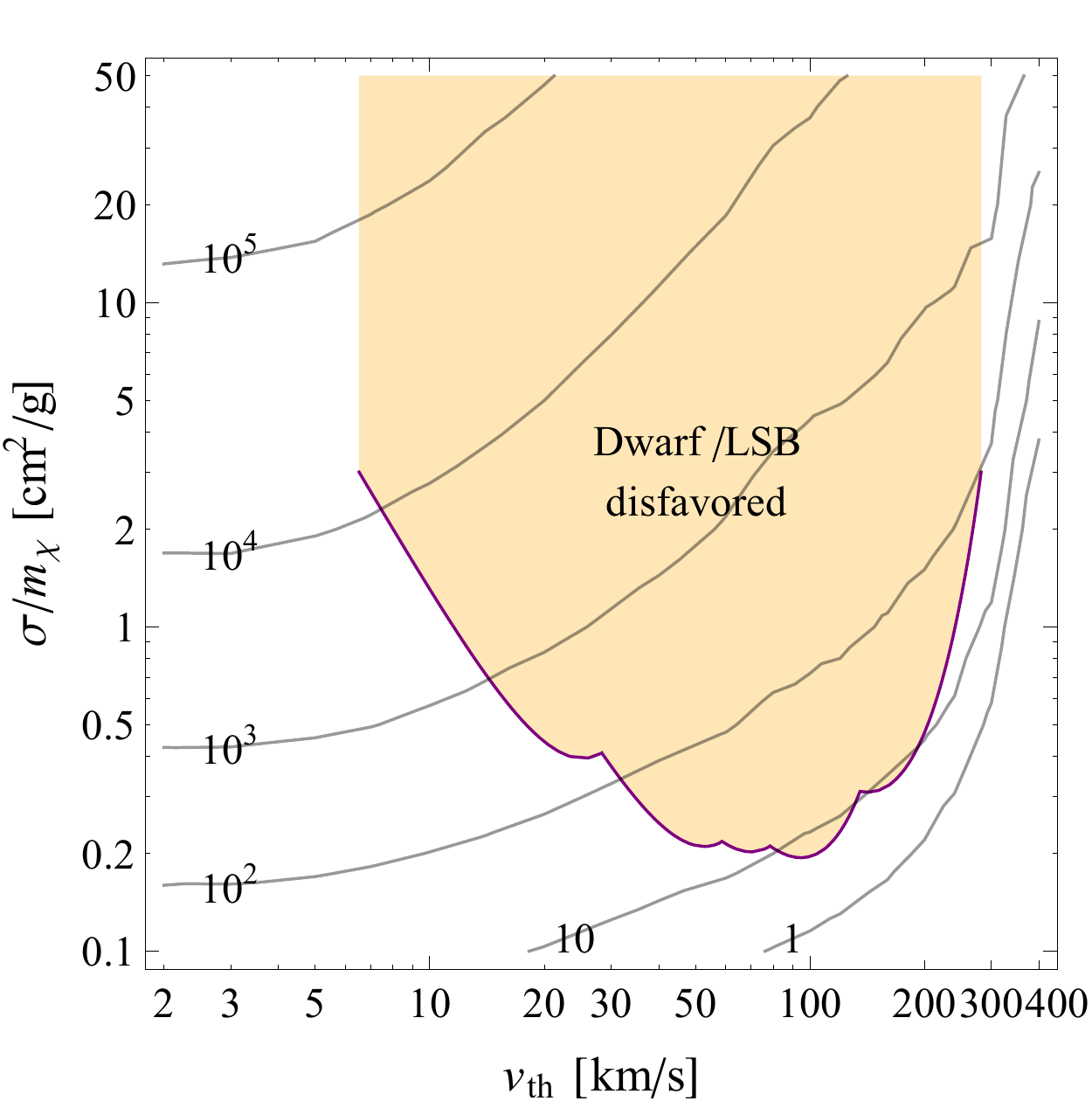}
\caption{The contour of the DDM average density enhancement factors \emph{from the sunk component only}, compared to the DM galactic component. The sunk component is counted by Eq.~\ref{eq:sunk} at a minimal semi-major axis of $a_\text{min}=1.455~R_\odot$, corresponding to our minimally tracked energy state $E_\text{min}/m_\chi=-2^{16}~(\text{km}/\text{s})^2$ in Eq.~\ref{eq:Erange}, as the boundary between the bound (orbited) component and the sunk component. Then the enhancement factor is calculated as $m_\chi N_\text{sunk}/(\rho_\chi\frac{4}{3}\pi a_\text{min}^3)$. Note that further channels that cannot captured in our calculation may further significantly affect the enhancement, see the text. We also plot the constraint from~\cite{Essig:2018pzq} for reference, which differs from our scenario by always assuming an elastic cross section of $3~\text{cm}^2/\text{g}$.}
\label{fig:Enhance}
\end{figure}
In such definition our results in Fig.~\ref{fig:Enhance} shows the possibility of a great enhancement to the local DM density, due to the sunk DDM component which is inside or extremely close to the Sun. In general we can see that the sunk component will increase in the direction of small $v_\text{th}$ as well as large $\sigma/m_\chi$, and the average density can be several orders larger than the local DM density of the galactic component.

We warn the reader that this is far from a satisfying accurate estimation, but actually only the numerical guess we can reach in our approach. If we can manage to include the sunk component in the initial state in our calculation, by the same enumeration we should have the additional sunk-sunk to sunk-sunk process (which just redistribute the sunk component and is not directly interesting in counting the total sunk DDM number), the sunk-free initial state to free-free, bound-bound or sunk-sunk processes, and the sunk-bound initial state to bound-bound or sunk-sunk processes. They are all beyond our current calculation. Among them, only the ones with the sunk-free initial state can change the total particle number of the bounded halo, while the other processes do not. The sunk-free to free-free process is in the direction to reduce the sunk DDM number, but it should be limited from the point of view of the available initial free DDM phase space, which should be only an energetic corner. On the other hand, the sunk-free to bound-bound process and the sunk-free to sunk-sunk process will further accrete the free DDM particles onto the bounded halo and increase the sunk component eventually. Although not directly contributing DDM particles from connection to the free component, the sunk-bound ones and the sunk-sunk ones are helping the deceleration of the bounded halo particles and the accretion, which is also in the direction of further increasing the sunk component. Since in the last two processes the second bounded DDM particle for the first sunk DDM to scatter has already a density much larger than the free galactic component density in the dominant central region (from Fig.~\ref{fig:forRforT} and \ref{fig:Enhance}), we can expect the effect of further increasing the sunk component will dominate over the effect of reducing the sunk component by the sunk-free to free-free process. Namely we expect the true DDM halo density inside or in the close neighborhood of the Sun is even more enhanced than what we guess from the Eq.~\ref{eq:sunk} and show in Fig.~\ref{fig:Enhance}. Such processes have the sunk component in the initial state, therefore the further increase behavior can be exponential, and the final DDM density can be further boosted by several orders.

Moreover, till now we have completely neglected the dark radiation which is assumed as promptly going away after the up-scattering. For low DDM density it should be fine, however, if the DDM dissipative scattering happens at a high enough rate, the dark radiation will exert a pressure on the DDM particles, expelling them from the center of the Sun. And this effect will presumably depend on other parameters beyond the cross section and threshold velocity, which further complicates the determination of the very inner region of the halo.

In Fig.~\ref{fig:Enhance} we have also plotted the constraint from Ref.~\cite{Essig:2018pzq}. Note the model assumption therein differs from ours by always assuming an elastic cross section of $3~\text{cm}^2/\text{g}$. We can find an area avoiding the constraint, that still has large enhancement and will be particularly interesting for DM indirect detection. One may generally think that the DDM accretion at the galaxy scale to speed up structure formation such as the core collapse and the accretion at the solar scale to form the compact halo are quite similar, so they should be optimized at exactly the same $\sigma/m_\chi$ and $v_\text{th}$ combination, and the most optimistic parameter for the solar halo enhancement should have been ruled out at the galaxy scale if the structure observed there is not that extreme. But in fact the solar case has the Sun as the external gravitational source which is many orders larger in providing gravity, and the accretion is not purely self-driven as the galaxy case. At the galaxy scale the optimistic $v_\text{th}$ for core collapse should be close to the characteristic circular velocity, and even on the low mass tail of the dwarf galaxy distribution such characteristic circular velocity is in practical bounded from below. On the other hand, the $v_\text{th}\to0$ limit is actually the elastic SIDM limit, on the solar scale side eventually the free-free-in process as the source should vanish if the DM self-scattering turns into purely elastic, and consequently all the accrete rate. But numerically we have not yet achieved that far in Fig.~\ref{fig:Enhance}. With the Sun providing the external driving of the accretion, arguably the region left to the exclusion in Fig.~\ref{fig:Enhance} or its expected update in the future is a viable parameter region to achieve significant density enhancement.

\subsection{Comment on Other Cases}
We have also performed similar calculations with the earth as the central gravitational source. Except for adopting different values for the earth, the other key difference is that there is also the effect of the solar gravitational field, which is equivalent to an extra gap of $-GMm_\chi/\text{AU}$ between the local infinity to the free galactic component in energy. We found that with a reasonable cross section as currently taken, the resultant halo will always be several orders below the galactic component, so it is less interesting. Given the values of the standard halo model, similar calculation can be performed to other nearby celestial bodies, \emph{e.g.}, a black hole. In such cases the validity of the elliptical orbit approximation can extend far deeper than that in the solar case, even if the black hole has a similar mass and consequently a similar DDM halo outskirt.

On the other hand, on the model side, the constant cross section is easily substituted by a velocity dependent cross section such as $\sigma/m_\chi\propto v^{-2}$ (Sommerfeld enhancement) or $v^{-4}$ (Coulomb scattering), up to model preferences.

\section{Indirect Detection Implications}\label{sec:6}

\subsection{Gamma-Ray Emission from the Sun}
The Sun is known to be a strong $\gamma$-ray source due to the hadronic interactions between cosmic ray nuclei and the solar atmosphere, as well as the inverse Compton scattering (ICS) between cosmic ray electrons and the sunlight~\cite{Seckel:1991ffa,Moskalenko:2006ta}. The Fermi-LAT observations do reveal a disk component as expected from the solar atmospheric interactions and a more extended component from the ICS interactions~\cite{Abdo:2011xn}. The measured $\gamma$-ray fluxes and energy spectra are different from the naive expectation based on the cosmic ray spectra, implying possible significant roles of the magnetic fields around the Sun~\cite{Abdo:2011xn,Ng:2015gya,Zhou:2016ljf}. More detailed analyses of the Fermi-LAT data revealed even more complicated temporal and spectral features of the solar $\gamma$-rays~\cite{Linden:2018exo,Tang:2018wqp}. Surprisingly, a statistically significant ``dip'' feature has been detected at around $30\sim50~\text{GeV}$ with a significance higher than $5\sigma$, which is lack of a reasonable explanation yet~\cite{Tang:2018wqp}.

Motivated by the possibility of compact DDM halo with significantly enhanced density, we consider the DM origin of the dip structure. We assume that the DDM annihilation products are $e^+e^-$ leptons. Then there are three relevant major contributing mechanisms to the observed $\gamma$-ray spectrum. The internal bremsstrahlung emission associated with the charged lepton final state (the final state radiation or FSR) and the ICS emission from $e^+e^-$ scattering off the sunlight contribute to the $\gamma$-ray spectrum mainly below the ``dip''. Since the magnetic field around the Sun is strong enough to confine charged particles below $\sim\text{TeV}$ energies~\cite{Zhou:2016ljf}, in estimation we adopt the {\it in situ} cooling approximation. On the other hand, in order to give the upper half of the ``dip'', we further consider the contribution from the virtual internal bremsstrahlung (VIB) process via the exchange of a virtual charged particle~\cite{Bringmann:2007nk,Bringmann:2012vr}. The spectrum of the VIB emission is very hard, which can mimic the monochromatic $\gamma$-ray emission given finite energy resolution of the detector~\cite{Bringmann:2012vr}. Its amplitude depends on the mass splitting between the mediator $\eta$ and the DM $\chi$, which is characterized by $\mu\equiv (m_\eta/m_\chi)^2$ with the best fit of $1.5$. Matching the ``dip'' position, the DDM mass should be $90~\text{GeV}$. And the velocity weighted average annihilation cross section is adopted $\langle\sigma v\rangle=3\times10^{26}~\text{cm}^3\text{s}^{-1}$, namely the value consistent with the thermal freeze-out mechanism. For the background contribution, we adopt the simulation results given in Ref.~\cite{Mazziotta:2020uey} with the potential field source surface~\cite{PFSS:1969} magnetic field model, together with an enhancement of the BIFROST model near the Sun~\cite{Gudiksen:2011cx}.

In all, the observational spectrum can be reasonably well fitted, only given an extremely large DDM density enhancement around the Sun. As the key, the DDM halo profile relevant to the phenomena is the region $r\gtrsim R_\odot$. For example, if the DDM density profile outside the Sun is adopted to be a power-law form $\propto r^{-1.6}$ as given in Fig.~\ref{fig:forRforT}, then the best fit DDM density at the Sun's surface is $\rho_\chi(R_\odot)=3.8\times10^7\rho_\chi$ with $\rho_\chi=0.39~\text{GeV}/\text{cm}^3$ as mentioned earlier; and (since we also do not know the profile of the DDM halo) if we switch the profile to a power law continuously connecting $\rho_\chi(R_\odot)$ and $\rho_\chi(2R_\odot)\approx\rho_\chi$ (assumed), in order to get the same spectrum we need roughly $\rho_\chi(R_\odot)=2.2\times10^8\rho_\chi$. This surface DDM density is still several orders larger than what we can get from our incomplete calculation such as in Fig.~\ref{fig:Enhance}, so we consider this fitting to be premature and not a rigorous interpretation for the $\gamma$-ray spectrum dip. However, as we commented earlier, we expect the true DDM halo density inside or in the close neighborhood of the Sun is even more enhanced (potentially by several orders) than what we guessed there, so we still consider this effort as a reasonable and interesting step towards a completely satisfying interpretation.

\subsection{Other Constraints}
For canonical DM density profiles (\emph{e.g.}, the Navarro-Frenk-White distribution~\cite{Navarro:1996gj} or NFW) and $m_\chi\sim100$~GeV, the Fermi-LAT $\gamma$-ray observations constrain the annihilation cross section to the $e^+e^-$ channel (or the $\mu^+\mu^-$ channel which gives quite similar constraints to that of the $e^+e^-$ channel) to be a few times of $\sim10^{-26}$~cm$^3$~s$^{-1}$~\cite{Huang:2012yf} and $\sim10^{-25}$~cm$^{-3}$~s$^{-1}$~\cite{Ackermann:2015zua}, from the Galactic center region and the combination of a population of dwarf spheroidal galaxies, respectively\footnote{The PLANCK observations of the cosmic microwave background anisotropies give similar constraints~\cite{Ade:2015xua}}. They are all consistent with the adopted value $\langle\sigma v\rangle=3\times10^{26}~\text{cm}^3\text{s}^{-1}$. However, in our DDM model the astrophysical profile of the DM halo should be different from the NFW one, and the accretion of the DDM in the MW center or the dwarf galaxies may also enhance the density distributions at these places, and change the constraints. Note that in reaching such a significant solar DM density enhancement, the favorite parameter region has a small $v_\text{th}$ from the implication of Fig.~\ref{fig:Enhance}, which is close to the $v_\text{th}\to0$ conventional elastic SIDM case as mentioned earlier. Pure SIDM simulation gives halo structure characterized by a constant density core in its central region, rather than the NFW cuspy proportional to $r^{-1}$~\cite{Vogelsberger:2012ku,Rocha:2012jg}. In such cases in the center of the MW or the dwarf galaxies, DM can be in fact less concentrated as their canonical DM density profile counterparts, and the above DM annihilation cross section bound should even be relaxed instead. On the other hand, the possibility that the DM concentration is more cuspy than its NFW counterpart can be indeed achieved. In particular in the specific MW satellite galaxy Draco, as shown in Ref.~\cite{Sameie:2019zfo}, an $\mathcal{O}(10)~\text{cm}^2/\text{g}$ purely elastic SIDM model with the MW tidal effect will lead to a core collapse configuration, which is indeed consistent with the observation. However, as mentioned earlier, the DM annihilation cross section constraint from Draco alone is not so strong. Simply put, we believe that the DM annihilation cross (as well as other parameters) adopted earlier in our calculation is consistent with the other observational constraints.

\section{Conclusion}\label{sec:7}

In this paper working in the dissipative self-interacting dark matter (DM) model, we have pointed out the possible existence of a compact DM halo around the Sun and other celestial bodies. We have defined a scheme of approximations to enable such calculation, and develop the Boltzmann equation set based on the partition function of the elliptical orbits characterized by its energy and angular momentum. Then we numerically solve the Boltzmann equation set to demonstrate the existence of the DM halo. Our results show that the DM density enhancement can be several orders in a compact region centered around the Sun. As an application, we use such possibility to study the DM origin of the unexpected ``dip'' structure in the observed solar $\gamma$-ray spectrum, which is reproduced by the inverse Compton scattering emission on the low energy side and the virtual internal bremsstrahlung mechanism on the high energy side, respectively.

Our results are incomplete and limited by the validity of the elliptical orbit approximation. Therefore we cannot directly sample all the bounded DM particles, but have to truncate at the lower boundary of the elliptical orbits, which renders the ``sunk'' component calculation, the inner region halo determination as well as the attempt of the DM origin of the solar $\gamma$-ray ``dip'' structure not rigorous. One may consider to completely numerically study the ``sunk'' orbits, but we reserve this for a future work.

{\it Acknowledgments}:\\
The author is grateful to useful discussion with Kenny Ng, Hai-bo Yu, Qiang Yuan and Yi-Ming Zhong. The credit towards interpreting the solar gamma-ray spectrum dip should totally go to Qiang Yuan. The author acknowledge the computational facilities of the chepfarm cluster at Tsinghua University.

\bibliography{LocalHalo}

\end{document}